\documentclass[english]{IEEEtran}
\usepackage[T1]{fontenc}
\usepackage[latin9]{inputenc}
\usepackage{float}
\usepackage{amsmath}
\usepackage{amssymb}
\usepackage{graphicx}

\makeatletter

\floatstyle{ruled}
\newfloat{algorithm}{tbp}{loa}
\providecommand{\algorithmname}{Algorithm}
\floatname{algorithm}{\protect\algorithmname}

\makeatother

\usepackage{babel}
\begin{document}
\title{Reliable Detection for Spatial Modulation Systems}
\author{Ibrahim Al-Nahhal$^{\ast}$, Octavia A. Dobre$^{\ast}$ and Salama
Ikki$^{\ddagger}$\\
$^{\ast}$Faculty of Engineering and Applied Science, Memorial University,
St. John\textquoteright s, Canada\\
$^{\ddagger}$Department of Electrical Engineering, Lakehead University,
Thunder Bay, Canada\\
\vspace{-5mm}
}
\maketitle
\begin{abstract}
Spatial modulation (SM) is a promising multiple-input multiple-output
system used to increase spectral efficiency. The maximum likelihood
(ML) decoder jointly detects the transmitted SM symbol, which is of
high complexity. In this paper, a novel reliable sphere decoder (RSD)
algorithm based on tree-search is proposed for the SM system. The
basic idea of the proposed RSD algorithm is to reduce the size of
the tree-search, and then, a smart searching method inside the reduced
tree-search is performed to find the solution. The proposed RSD algorithm
provides a significant reduction in decoding complexity compared to
the ML decoder and existent decoders as well. Moreover, the RSD algorithm
provides a flexible trade-off between the bit error rate (BER) performance
and decoding complexity, so as to be reliable for a wide range of
practical hardware implementations. The BER performance and decoding
complexity analysis for the RSD algorithm are studied, and Monte Carlo
simulations are then provided to demonstrate the findings.
\end{abstract}


\section{Introduction}

\def\figurename{Fig.}
\def\tablename{TABLE}

Spatial modulation (SM) is a promising technique {[}\ref{Ref: Ch7_1}{]}
that has been recently applied to many of the emerging technologies
{[}\ref{Ref: Ch7_2}{]}, {[}\ref{Ref: Ch7_3}{]}. It overcomes the
inter-channel interference (ICI) problem that exists in multiple-input
multiple-output (MIMO) systems. The SM system completely eliminates
the ICI by delivering a phase-shift-keying (PSK) or quadrature amplitude
modulation (QAM) symbol from only one transmit antenna at a time.
A part of the input bit-stream determines an active transmit antenna,
while the rest determines the PSK/QAM symbol to be delivered from
the activated antenna {[}\ref{Ref: Ch7_4}{]}, {[}\ref{Ref: Ch7_5}{]}.
At the receiver, the maximum-likelihood (ML) decoder is applied to
obtain the optimum bit error rate (BER) at the expense of the decoding
complexity {[}\ref{Ref: Ch7_6}{]}.

Several low-complexity decoding algorithms have been recently proposed
in {[}\ref{Ref: Ch7_7}{]}-{[}\ref{Ref: Ch7_12}{]} to reduce the
high decoding complexity of the ML decoder. In {[}\ref{Ref: Ch7_7}{]}
and {[}\ref{Ref: Ch7_8}{]}, the sphere decoder (SD) concept is utilized
to reduce the decoding complexity without sacrificing the optimum
BER performance. A low-complexity decoding algorithm has been proposed
in {[}\ref{Ref: Ch7_9}{]} and extended in {[}\ref{Ref: Ch7_10}{]}
by exploiting a smart searching algorithm in the tree-search to obtain
the optimum BER performance. The authors in {[}\ref{Ref: Ch7_11}{]}
and {[}\ref{Ref: Ch7_12}{]} proposed low-complexity decoders by sacrificing
the optimality of the BER performance. The existing SD algorithms
suffer from a lack of reliability when it comes to fitting the practical
hardware implementation requirements. In other words, the existing
algorithms do not provide a suitable trade-off between the BER performance
and decoding complexity.

This paper proposes a novel and reliable SD (RSD) algorithm that provides
an advantageous arrangement between the BER performance and decoding
complexity. Besides, the proposed RSD algorithm can achieve the optimum
BER performance with a significant reduction in the decoding complexity
compared to the ML decoder and the existing algorithms as well. The
analytical BER analysis and expected decoding complexity of the proposed
algorithm are provided and confirmed through Monte Carlo simulations.

\section{System Model}

Consider an $N_{r}\times N_{t}$ SM-MIMO system, where $N_{r}$ and
$N_{t}$ represent the number of transmit and receive antennas, respectively.
SM delivers $\text{log}_{2}(N_{t}M)$ bit per channel use, where $M$
is the modulation order of the QAM constellation. The input bit-stream
is split into two groups: the first $\text{log}_{2}(N_{t})$ bits
select the active antenna, while the second $\text{log}_{2}(M)$ bits
determine the QAM symbol to be transmitted, $s_{t}\in\{s_{1},\cdots,s_{M}\}$.
The SM transmitted message, $\boldsymbol{x}_{t}$, is equal to $\boldsymbol{h}_{t}s_{t}$,
where $\boldsymbol{h}_{t}$ is a vector of the Rayleigh fading channel
coefficients with entries distributed as $\mathcal{C}\mathcal{N}(0,1)$
and drawn from the channel matrix, $\boldsymbol{H}\in\mathbb{C}^{N_{r}\times N_{t}}$.
The received signal is

\vspace{-4mm}

\begin{equation}
\boldsymbol{y}=\boldsymbol{x}_{t}+\boldsymbol{g},\label{eq: y_SM}
\end{equation}

\noindent where $\boldsymbol{g}\sim\mathcal{C}\mathcal{N}(0,\sigma_{g}^{2})$
is the vector of additive white Gaussian noise (AWGN) samples.

At the receiver side, the ML decoder estimates the transmitted SM
message, $\hat{\boldsymbol{x}}_{\text{ML}}$, as {[}\ref{Ref: Ch7_6}{]}

\vspace{-4mm}

\begin{equation}
\hat{\boldsymbol{x}}_{\text{ML}}\hspace{-0.9mm}=\hspace{-0.9mm}\hspace{-0.9mm}\hspace{-1.2mm}\hspace{-1.5mm}\underset{\boldsymbol{x}_{j}|j=1,\cdots\hspace{-0.9mm},MN_{t}}{\text{arg\,min}}\hspace{-1.5mm}\left\Vert \boldsymbol{y}-\boldsymbol{x}_{j}\right\Vert ^{2}\hspace{-0.9mm}=\hspace{-0.9mm}\hspace{-0.9mm}\hspace{-1.2mm}\hspace{-1.5mm}\underset{\boldsymbol{x}_{j}|j=1,\cdots,MN_{t}}{\text{arg\,min}}\sum_{i=1}^{N_{r}}\left|y_{i}-x_{i,j}\right|^{2}\hspace{-0.9mm}.\label{eq: J_ML}
\end{equation}

\begin{figure}
\begin{centering}
\includegraphics[scale=0.35]{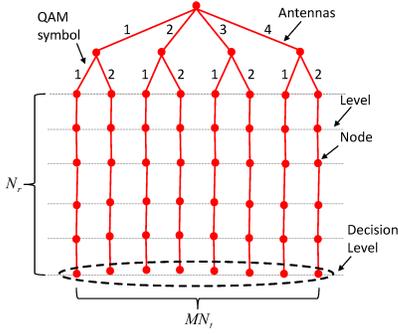}
\par\end{centering}
\caption{\label{fig:Ch_7_SM-tree-search-ML}{\small{}Tree-search of SM-MIMO
for $M=2$, $N_{t}=4$, and $N_{r}=6$.}}
\end{figure}

\noindent The tree-search structure {[}\ref{Ref: Ch7_8}{]}, {[}\ref{Ref: Ch7_10}{]}
can be used to represent (\ref{eq: J_ML}). The tree-search is a two-dimensional
structure with a size of $N_{r}\times MN_{t}$; the tree-search width
represents the $MN_{t}$ possibilities of the SM message called branches,
while its depth represents the $N_{r}$ levels of each possibility
of the SM message. Fig. \ref{fig:Ch_7_SM-tree-search-ML} shows a
tree-search example of the ML decoder for $M=2$, $N_{t}=4$, and
$N_{r}=6$. The accumulated distance metric vector of the $i$-th
level, $\boldsymbol{v}(i)\in\mathbb{R}^{1\times MN_{t}}$, is

\vspace{-4mm}

\begin{equation}
\boldsymbol{v}(i)\hspace{-0.9mm}=\hspace{-0.9mm}\left[\hspace{-1.5mm}\begin{array}{ccc}
\sum_{n=1}^{i}\left|y_{n}-x_{n,1}\right|^{2}\hspace{-1.5mm} & \hspace{-0.9mm}\cdots\hspace{-0.9mm} & \hspace{-1mm}\sum_{n=1}^{i}\left|y_{n}-x_{n,MN_{t}}\right|^{2}\end{array}\hspace{-1.5mm}\right].\label{eq: v_i}
\end{equation}

\noindent Typically, the last level of the tree-search is called the
\textit{decision level}. The ML decoder estimates $\hat{\boldsymbol{x}}_{\text{ML}}$
that corresponds to the minimum node in $\boldsymbol{v}(N_{r})$.
Note that the SM tree-search is quite different than the MIMO tree-search
{[}\ref{Ref: Ch7_13}{]}-{[}\ref{Ref: Ch7_15}{]}.

In this paper, the decoding complexity is defined as the total number
of nodes that should be visited in the tree-search to estimate the
transmitted SM message. Since the ML decoder visits all nodes in the
tree-search, its decoding complexity is $\Psi^{\text{ML}}=MN_{t}N_{r}$.

\noindent The complexity of the ML decoder consequently becomes extensive,
especially for higher SM-MIMO dimensions and/or QAM sizes. Several
works in the literature have been proposed to reduce the ML complexity,
which are based on tree-search and SD concepts. However, further complexity
reduction can still be achieved, as well as progress towards its reliability
to fit a wide range of hardware implementation.

\vspace{-2mm}

\section{\label{sec:Proposed-Fixed-Complexity-Algori}The Proposed RSD Algorithm}

The proposed RSD algorithm firstly reduces the size of the tree-search,
and then performs a smart searching method to reach the solution.
Let us define $\psi_{\text{col}}$ as the number of branches/SM message
possibilities that most likely contains the optimum solution. The
RSD algorithm performs its searching for the solution inside these
$\psi_{\text{col}}$ branches and stops at the $\psi_{\text{row}}$-th
level, where $1\leq\psi_{\text{row}}\leq N_{r}$ is the maximum number
of levels that can be visited by the RSD algorithm (i.e., the decision
level at $\psi_{\text{row}}$). It is worth noting that the flexibility
trade-off between the BER performance and complexity provided by the
RSD algorithm comes from changing the value of $\psi_{\text{row}}$
within the range of $1$ and $N_{r}$.

The steps of searching for the solution of the RSD algorithm inside
the reduced tree-search are as follows:

\textbf{\textit{Step 1:}} Expand all nodes of the first level, i.e.,
$\boldsymbol{v}(1)$ in (\ref{eq: v_i}).

\textbf{\textit{Step 2:}} Appropriately choose the smallest $\psi_{\text{col}}$
nodes that come from \textbf{\textit{Step 1}}. It should be noted
that the RSD algorithm searches for the solution inside the branches
that correspond to the smallest $\psi_{\text{col}}$ nodes. Consequently,
the RSD algorithm reduces the decoding complexity by at least $(MN_{t}-\psi_{\text{col}})\psi_{\text{row}}$
nodes. The vector of distance metric nodes in (\ref{eq: v_i}) yields

\vspace{-3mm}

\begin{equation}
\boldsymbol{v}(i)=\hspace{-0.9mm}\left[\hspace{-1.5mm}\begin{array}{ccccc}
v(i,1) & \cdots & \hspace{-0.9mm}v(i,j) & \hspace{-0.9mm}\hspace{-0.9mm}\cdots\hspace{-0.9mm} & \hspace{-0.9mm}v(i,\psi_{\text{col}})\end{array}\hspace{-1.5mm}\right],\label{eq: v_i_RSD}
\end{equation}

\noindent where $v(i,j)$ is the $j$-th node of level $i$, and given
by

\vspace{-3mm}

\begin{equation}
v(i,j)=\sum_{n=1}^{i}\left|y_{n}-x_{n,j}\right|^{2}.\label{eq: v(i,j)}
\end{equation}

\begin{algorithm}[t]
\begin{itemize}
\item \textbf{Input} $\boldsymbol{H}$, $M$, and $N_{t}$;
\item \textbf{Compute} $\boldsymbol{v}(1)$ in (\ref{eq: v_i});
\item \textbf{Choose }$\psi_{\text{col}}$ empirically, based on $M$ and
$N_{t}$ to most likely include the optimum BER performance;
\item \textbf{Store} the branches indices that corresponding to the smallest
$\psi_{\text{col}}$ node of $\boldsymbol{v}(1)$ into $\varXi_{\psi_{\text{col}}}$;
\item \textbf{Choose} $\psi_{\text{row}}$ based on the system requirements
from the BER and complexity points of views;
\item \textbf{Define} $\text{Len}(j)$ as the length of the $j$-th branch
and initiate it with one for $\forall j$;
\end{itemize}
{\small{}~~~~~1: }\textbf{\textit{While$\,\,\,n<\psi_{\text{row}}\psi_{\text{col}}$,
do}}

{\small{}~~~~~2:}\textbf{\small{} ~~~}\textbf{Find}\textbf{\small{}
}$j_{\text{min}}$ that solves $\underset{\begin{array}{c}
j\in\varXi_{\psi_{\text{col}}}\\
i_{\text{min}}\in\{1,\cdots,\psi_{\text{row}}\}
\end{array}}{\text{arg\,min}}\hspace{-3mm}\left\{ \boldsymbol{v}(i_{\text{min}})\right\} $;

{\small{}~~~~~3:}\textbf{\small{} ~~~}\textbf{Update $i_{\text{min}}$
}as the level that corresponding to $j_{\text{min}}$;

{\small{}~~~~~4: ~~~}\textbf{\textit{ if }}$\text{Len}(j_{\text{min}})==\psi_{\text{row}}$

{\small{}~~~~~5:}\textbf{\small{} ~~~~~}\textbf{ break}\textbf{\small{}
}and end the algorithm;

{\small{}~~~~~6: ~~~}\textbf{\textit{ else}}

{\small{}~~~~~7:}\textbf{\small{} ~~~~~}\textbf{ Expand
$v(i_{\text{min}},j_{\text{min}})\leftarrow v(i_{\text{min}}+1,j_{\text{min}})$};

{\small{}~~~~~8:}\textbf{\small{} ~~~~~}\textbf{ Update
$\boldsymbol{v}(i_{\text{min}})$ }based on\textbf{ $v(i_{\text{min}},j_{\text{min}})$};

{\small{}~~~~~9: ~~~}\textbf{\textit{ end if}}

{\small{}~~~~10: }\textbf{\small{}~~}\textbf{ Set $n\leftarrow n+1$};

{\small{}~~~~11: }\textbf{\textit{end While}}
\begin{itemize}
\item \textbf{Output}{\small{} }$\hat{\boldsymbol{x}}_{\text{RSD}}=\underset{\begin{array}{c}
j\in\varXi_{\psi_{\text{col}}}\end{array}}{\text{arg\,min}}\left\{ \boldsymbol{v}(\psi_{\text{row}})\right\} $.
\end{itemize}
\caption{\label{alg: RSD}The proposed RSD algorithm pseudo-code.}
\end{algorithm}

\textbf{\textit{Step 3:}} Perform a single expansion to the minimum
node in (\ref{eq: v_i_RSD}).

\textbf{\textit{Step 4:}} Check if the expanded node from \textbf{\textit{Step
3}} still has a minimum value among the rest of $\psi_{\text{col}}$
nodes or not. If yes, perform another single expansion to that node.
If no, find the new minimum node and expand it once.

\textbf{\textit{Step 5:}} Repeat \textbf{\textit{Step 3}} and \textbf{\textit{Step
4}} until the RSD algorithm obtains the minimum node at a branch with
a length of $\psi_{\text{row}}$.

\textbf{\textit{Step 6:}} Find the index corresponding to the node
that comes from \textbf{\textit{Step 5}}, and declare it as the solution
of the RSD algorithm. The estimated SM message using the RSD algorithm,
$\hat{\boldsymbol{x}}_{\text{RSD}}$, can be given as

\vspace{-3mm}

\[
\hat{\boldsymbol{x}}_{\text{RSD}}=\underset{\begin{array}{c}
\boldsymbol{x}_{j}|j\in\varXi_{\psi_{\text{col}}}\end{array}}{\text{arg\,min}}\sum_{i=1}^{\psi_{\text{row}}}\left|y_{i}-x_{i,j}\right|^{2}
\]

\vspace{-3mm}

\begin{equation}
=\underset{\begin{array}{c}
j\in\varXi_{\psi_{\text{col}}}\end{array}}{\text{arg\,min}}\left\{ \boldsymbol{v}(\psi_{\text{row}})\right\} ,\label{eq: J_CRB}
\end{equation}

\noindent where $\varXi_{\psi_{\text{col}}}$ denotes the set of branch
indices that corresponds to the smallest $\psi_{\text{col}}$ metric
node values of $\boldsymbol{v}(1)$ (i.e., the first level at $i=1$
in (\ref{eq: v_i})). The RSD algorithm is summarized in Algorithm
\ref{alg: RSD}.

\vspace{-3mm}

\section{Theoretical Analysis}

The RSD algorithm provides the optimum BER performance with a significant
reduction in the decoding complexity. In addition, by changing the
value of $\psi_{\text{row}}$, a flexible trade-off between the BER
performance and decoding complexity can be obtained to fit a wide
range of hardware implementation. In this section, the BER performance
and expected complexity are considered random variables, and their
approximate expressions are derived using the probability theory.

\vspace{-1.5mm}

\subsection{\label{subsec:BER-Upper-Bound}BER Upper Bound Analysis}

The general expression for the upper bound of the ML BER for SM is
{[}\ref{Ref: Ch7_6}{]}, {[}\ref{Ref: Ch7_16}{]}

\vspace{-4mm}

\begin{equation}
\text{BER}^{\text{ML}}\leq\sum_{j=1}^{MN_{t}}\sum_{\hat{j}=1}^{MN_{t}}\frac{\delta(\boldsymbol{x}_{j},\hat{\boldsymbol{x}}_{\hat{j}})\mathcal{\mathbb{E}}\left\{ \mathbb{P}\text{r}^{\text{ML}}\left(\boldsymbol{x}_{j}\rightarrow\hat{\boldsymbol{x}}_{\hat{j}}\right)\right\} }{MN_{t}\text{log}_{2}(MN_{t})},\label{eq: UB BER general}
\end{equation}

\vspace{-1.5mm}

\noindent where $\mathbb{P}\text{r}^{\text{ML}}(\boldsymbol{x}_{j}\rightarrow\hat{\boldsymbol{x}}_{\hat{j}})$
is the pairwise error probability (PEP) of the ML algorithm, $\mathbb{P}\text{r}(\centerdot)$
denotes the probability of an event, $\mathcal{\mathbb{E}}\left\{ \cdot\right\} $
represents the expectation operation, and $\delta(\boldsymbol{x}_{j},\hat{\boldsymbol{x}}_{\hat{j}})$
denotes the Hamming distance which measures the number of bits in
error between $\boldsymbol{x}_{j}$ and $\hat{\boldsymbol{x}}_{\hat{j}}$.

Since the RSD algorithm performs the search inside a portion of the
tree-search with a size of $\psi_{\text{row}}\times\psi_{\text{col}}$,
the optimal solution may not be included in that portion of the tree-search.
Thus, the PEP in (\ref{eq: UB BER general}) for the RSD algorithm
can be written as

\vspace{-3mm}

\begin{equation}
\mathbb{P}\text{r}^{\text{RSD}}\hspace{-1.1mm}\left(\hspace{-1.1mm}\boldsymbol{x}_{j}\rightarrow\hat{\boldsymbol{x}}_{\hat{j}}\hspace{-1.1mm}\right)\hspace{-0.9mm}=\hspace{-0.9mm}\mathbb{P}\text{r}\left(\hat{\boldsymbol{x}}_{\text{opt}}\hspace{-0.9mm}\neq\hspace{-0.6mm}\boldsymbol{x}_{t}|\hat{\boldsymbol{x}}_{\text{opt}}\hspace{-0.6mm}\in\hspace{-0.6mm}\varXi_{\psi_{\text{col}}}\right)+\mathbb{P}\text{r}\left(\hat{\boldsymbol{x}}_{\text{opt}}\hspace{-0.6mm}\notin\hspace{-0.6mm}\varXi_{\psi_{\text{col}}}\right),\label{eq: PEP RSD}
\end{equation}

\noindent where $\hat{\boldsymbol{x}}_{\text{opt}}$ is the optimal
solution. The conditional probability in (\ref{eq: PEP RSD}) contains
two independent events. The expected value of (\ref{eq: PEP RSD})
can consequently be written as

\vspace{-3mm}

\begin{equation}
\mathcal{\mathbb{E}}\left\{ \hspace{-0.6mm}\mathbb{P}\text{r}^{\text{RSD}}\hspace{-1.1mm}\left(\hspace{-1.1mm}\boldsymbol{x}_{j}\rightarrow\hat{\boldsymbol{x}}_{\hat{j}}\hspace{-1.1mm}\right)\hspace{-1.1mm}\right\} \hspace{-1.1mm}=\hspace{-0.9mm}\underset{\text{Term 1}}{\underbrace{\mathcal{\mathbb{E}}\left\{ \mathbb{P}\text{r}\left(\hat{\boldsymbol{x}}_{\text{opt}}\hspace{-0.9mm}\neq\hspace{-0.6mm}\boldsymbol{x}_{t}\right)\right\} }}+\underset{\text{Term 2}}{\underbrace{\mathcal{\mathbb{E}}\left\{ \mathbb{P}\text{r}\left(\hat{\boldsymbol{x}}_{\text{opt}}\hspace{-0.6mm}\notin\hspace{-0.6mm}\varXi_{\psi_{\text{col}}}\right)\right\} }}.\label{eq: PEP RSD final}
\end{equation}

\noindent Term 1 in (\ref{eq: PEP RSD final}) can be written as in
{[}\ref{Ref: Ch7_6}{]}, {[}\ref{Ref: Ch7_16}{]}

\vspace{-3mm}

\begin{equation}
\mathcal{\mathbb{E}}\left\{ \mathbb{P}\text{r}\left(\hat{\boldsymbol{x}}_{\text{opt}}\hspace{-0.9mm}\neq\hspace{-0.6mm}\boldsymbol{x}_{t}\right)\right\} \hspace{-0.9mm}=\hspace{-0.9mm}\mu_{j,\hat{j}}^{\psi_{\text{row}}}\hspace{-1mm}\hspace{-0.9mm}\sum_{k=0}^{\psi_{\text{row}}-1}\hspace{-1mm}\hspace{-0.9mm}\left(\hspace{-2mm}\begin{array}{c}
\psi_{\text{row}}-\hspace{-0.9mm}1\hspace{-0.9mm}+\hspace{-0.9mm}k\\
k
\end{array}\hspace{-2mm}\right)\hspace{-0.9mm}(1-\mu_{j,\hat{j}})^{k},\label{eq: PEP_ML}
\end{equation}

\noindent with

\vspace{-3mm}

\begin{equation}
\mu_{j,\hat{j}}\hspace{-0.9mm}=0.5\hspace{-0.9mm}\left(\hspace{-0.9mm}\hspace{-0.5mm}1-\hspace{-0.9mm}\sqrt{\hspace{-0.9mm}\frac{\sigma_{j,\hat{j}}^{2}}{1\hspace{-0.9mm}+\hspace{-0.9mm}\sigma_{j,\hat{j}}^{2}}}\right)\hspace{-0.9mm},\,\,\,\,\,\sigma_{j,\hat{j}}^{2}=\hspace{-0.9mm}\frac{\rho(|s(j)|^{2}\hspace{-0.9mm}+\hspace{-0.9mm}|s(\hat{j})|^{2})}{4},\label{eq: mu and sigma}
\end{equation}

\noindent where $\rho$ is the average signal to noise ratio (SNR),
and $s(j)$ is the QAM symbol of the $j$-th SM transmitted message.
Hence, for the RSD algorithm, (\ref{eq: UB BER general}) can be written
as

\vspace{-3mm}

\[
\text{BER}^{\text{RSD}}\leq\sum_{j=1}^{\psi_{\text{col}}}\sum_{\hat{j}=1}^{\psi_{\text{col}}}\frac{\delta(\boldsymbol{x}_{j},\hat{\boldsymbol{x}}_{\hat{j}})}{MN_{t}\text{log}_{2}(MN_{t})}\times\hspace{3cm}
\]

\vspace{-4mm}

\begin{equation}
\left[\left[\mu_{j,\hat{j}}^{\psi_{\text{row}}}\hspace{-2.1mm}\sum_{k=0}^{\psi_{\text{row}}-1}\hspace{-2mm}\left(\hspace{-2.2mm}\begin{array}{c}
\psi_{\text{row}}-\hspace{-0.9mm}1\hspace{-0.9mm}+\hspace{-0.9mm}k\\
k
\end{array}\hspace{-2.2mm}\right)\hspace{-1mm}(1-\mu_{j,\hat{j}})^{k}\hspace{-0.9mm}\right]\hspace{-1mm}+\hspace{-0.9mm}\mathcal{\mathbb{E}}\left\{ \mathbb{P}\text{r}\left(\hat{\boldsymbol{x}}_{\text{opt}}\notin\varXi_{\psi_{\text{col}}}\right)\right\} \hspace{-1mm}\right]\hspace{-1mm}.\label{eq: UB BER RSD}
\end{equation}

Based on (\ref{eq: UB BER RSD}), the RSD algorithm provides a near
optimum BER performance when Term 2 in (\ref{eq: PEP RSD final})
tends to be zero (i.e., $\mathbb{P}\text{r}(\hat{\boldsymbol{x}}_{\text{opt}}\notin\varXi_{\psi_{\text{col}}})\approx0$);
this can be achieved by properly choosing $\psi_{\text{col}}$. In
this paper, $\psi_{\text{col}}$ is empirically chosen such that $\mathbb{P}\text{r}(\hat{\boldsymbol{x}}_{\text{opt}}\notin\varXi_{\psi_{\text{col}}})\approx0$.

\subsection{\label{subsec:Expected-Complexity-Analysis}Expected Complexity Analysis}

In this paper, the complexity of the RSD algorithm is measured by
the number of visited nodes in the tree-search needed to estimate
the solution. In general, the complexity of the SD algorithms is a
random variable. The general approximation for the expected SD complexity
is {[}\ref{Ref: Ch7_10}{]}

\vspace{-3mm}

\begin{equation}
\Psi^{\text{SD}}\approx MN_{t}+\sum_{j=1}^{MN_{t}}\sum_{i=1}^{N_{r}}\mathbb{P}\text{r}\left(v(i,j)\leq\zeta\left|\boldsymbol{x}_{t},\mathbf{H},\sigma_{g}^{2},\zeta\right.\right),\label{eq: C_SD _expected}
\end{equation}

\noindent where $\Psi^{\text{SD}}$ is the expected complexity of
an SD algorithm and $\zeta$ is the pruned radius (i.e., threshold)
of that algorithm. It should be noted that (\ref{eq: C_SD _expected})
represents the general expression and its solution depends on the
algorithm itself.

To find the conditional probability in (\ref{eq: C_SD _expected})
for the RSD algorithm, the distributions of $v(i,j)$ and $\zeta$
should be defined. From (\ref{eq: v(i,j)}), $v(i,j)$ has a non-central
chi-square distribution with $2i$ degrees of freedom. Thus, the closed-form
of the conditional probability in (\ref{eq: C_SD _expected}) is {[}\ref{Ref: Ch7_17},
(Ch. 2){]}

\vspace{-4mm}

\begin{equation}
\mathbb{P}\text{r}\left(v(i,j)\leq\zeta\left|\boldsymbol{x}_{t},\mathbf{H},\sigma_{g}^{2},\zeta\right.\right)=\hspace{-0.9mm}1-Q_{i}\hspace{-0.9mm}\left(\hspace{-0.9mm}\frac{\sqrt{2\gamma_{i,j}^{2}}}{\sigma_{g}}\,,\hspace{-0.9mm}\,\frac{\sqrt{2\zeta}}{\sigma_{g}}\hspace{-0.9mm}\right)\hspace{-0.9mm},\label{eq: Conditional_pr_rho}
\end{equation}

\vspace{-2.5mm}

\noindent where $\gamma_{i,j}^{2}=\sum_{n=1}^{i}|x_{n,t}-x_{n,j}|^{2}$,
$x_{n,j}$ is the $n$-th element of the $j$-th SM transmitted message,
and $Q_{i}(\centerdot,\centerdot)$ is the Marcum Q-function.

To remove the dependency of (\ref{eq: Conditional_pr_rho}) on $\zeta$,
the expectation operation should be applied for (\ref{eq: Conditional_pr_rho})
over the distribution of $\zeta$. For simplicity, let us assume that
the RSD algorithm most likely reaches the optimum solution. Thus,
the pruned radius can be given from $\zeta=\sum_{i=1}^{\psi_{\text{row}}}|g_{i}|^{2}$,
where $g_{i}$ denotes the $i$-th element of the AWGN vector in (\ref{eq: y_SM}).
It is worth noting that this simplified assumption of $\zeta$ is
especially for high SNR. The distribution of $\zeta$ is a central
chi-square with $2\psi_{\text{row}}$ degrees of freedom and its probability
density function, $f_{\zeta}(\zeta)$, is {[}\ref{Ref: Ch7_17}, (Ch.
2){]}

\vspace{-3.5mm}

\begin{equation}
f_{\zeta}(\zeta)=\frac{\left(\zeta\right)^{\psi_{\text{row}}-1}}{\sigma_{g}^{2\psi_{\text{row}}}\left(\psi_{\text{row}}-1\right)!}\,\text{exp}\left(\frac{-\zeta}{\sigma_{g}^{2}}\right).\label{eq:R_RSD pdf}
\end{equation}

\noindent Hence, (\ref{eq: Conditional_pr_rho}) yields

\vspace{-5mm}

\begin{equation}
\mathbb{P}\text{r}\left(v(i,j)\leq\zeta\left|\boldsymbol{x}_{t},\mathbf{H},\sigma_{g}^{2}\right.\right)\hspace{-0.7mm}=\hspace{-0.7mm}1-\hspace{-0.7mm}\int_{0}^{\infty}\hspace{-2mm}Q_{i}\hspace{-1mm}\left(\hspace{-1mm}\frac{\sqrt{2\gamma_{i,j}^{2}}}{\sigma_{g}}\,,\,\frac{\sqrt{2\zeta}}{\sigma_{g}}\hspace{-0.7mm}\right)\text{d}\zeta.\label{eq:RSD CDF independent}
\end{equation}

The closed-form expression of (\ref{eq:RSD CDF independent}) can
be given as {[}\ref{Ref: Ch7_18}{]}

\vspace{-3.5mm}

\[
\mathbb{P}\text{r}\left(v(i,j)\leq\zeta\left|\boldsymbol{x}_{t},\mathbf{H},\sigma_{g}^{2}\right.\right)=2^{i}\,\text{exp}\left(\frac{-\gamma_{i,j}^{2}}{\sigma_{g}^{2}}\right)\hspace{2cm}
\]

\vspace{-3.5mm}

\begin{equation}
\hspace{2.5cm}\times\sum_{n=0}^{\psi_{\text{row}}-1}\frac{\left(i\right)_{n}}{2^{n}\,n!}\,_{1}F_{1}\left(n+i;i;\frac{\gamma_{i,j}^{2}}{2\sigma_{g}^{2}}\right),\label{eq:RSD CDF independent_closed form}
\end{equation}

\noindent where $\left(i\right)_{n}$ represents the Pochhammer symbol
and $_{1}F_{1}$ is the Kummer hypergeometric function. Since the
RSD algorithm searches for the solution inside a reduced tree-search
with a size of $\psi_{\text{row}}\times\psi_{\text{col}}$, the approximation
of the expected complexity in (\ref{eq: C_SD _expected}) becomes

\[
\Psi^{\text{RSD}}\approx\psi_{\text{col}}+\sum_{j=1}^{\psi_{\text{col}}}\sum_{i=1}^{\psi_{\text{row}}}2^{i}\,\text{exp}\left(\frac{-\gamma_{i,j}^{2}}{\sigma_{g}^{2}}\right)\hspace{2.3cm}
\]

\begin{equation}
\hspace{2.5cm}\times\sum_{n=0}^{\psi_{\text{row}}-1}\frac{\left(i\right)_{n}}{2^{n}\,n!}\,_{1}F_{1}\left(n+i;i;\frac{\gamma_{i,j}^{2}}{2\sigma_{g}^{2}}\right),\label{eq: RSD _expected comp}
\end{equation}

\noindent where $\Psi^{\text{RSD}}$ is the expected complexity of
the RSD algorithm.

Alternatively, (\ref{eq:RSD CDF independent}) can be numerically
calculated using the Gauss--Laguerre quadrature {[}\ref{Ref: Ch7_19}{]}.
Thus, (\ref{eq: C_SD _expected}) becomes

\vspace{-4mm}

\[
\Psi^{\text{RSD}}\approx\psi_{\text{col}}\left(\psi_{\text{row}}+1\right)-\frac{1}{\left(\psi_{\text{row}}-1\right)!}\hspace{2.3cm}
\]

\vspace{-3mm}

\begin{equation}
\times\sum_{j=1}^{\psi_{\text{col}}}\sum_{i=1}^{\psi_{\text{row}}}\sum_{k=1}^{\beta}w_{k}\left(z_{k}\right)^{(\psi_{\text{row}}-1)}Q_{k}\hspace{-1mm}\left(\hspace{-1mm}\frac{\sqrt{2\gamma_{i,j}^{2}}}{\sigma_{g}}\,,\,\sqrt{2z_{k}}\hspace{-0.7mm}\right),\label{eq: RSD _expected comp numer}
\end{equation}

\noindent where $w_{k}$ and $z_{k}$ are given values based on the
order $\beta$, which is given from {[}\ref{Ref: Ch7_19}, (Table
25.9){]}. Note that (\ref{eq: RSD _expected comp numer}) provides
a close value to that in (\ref{eq: RSD _expected comp}) with considerably
lower execution time.

\section{Simulation Results}

In this section, the BER and ing complexity of the proposed RSD algorithm
are assessed and compared with optimum algorithms in literature, such
as {[}\ref{Ref: Ch7_7}{]}, {[}\ref{Ref: Ch7_8}{]}, and {[}\ref{Ref: Ch7_10}{]}.
Two SM-MIMO systems are considered; 16-QAM for $8\times8$ and $16\times16$
SM-MIMO, respectively. As mentioned before, $\psi_{\text{col}}$ is
empirically chosen to provide the optimum BER performance (i.e., $\mathbb{P}\text{r}(\hat{\boldsymbol{x}}_{\text{opt}}\notin\varXi_{\psi_{\text{col}}})\approx0$
in (\ref{eq: UB BER RSD})) at $\psi_{\text{row}}=N_{r}$, where $\psi_{\text{col}}=70$
and $180$ for the first and second SM-MIMO systems, respectively.
The proposed RSD algorithm is denoted by RSD-($\psi_{\text{row}}$,$\psi_{\text{col}}$)
to show the values of $\psi_{\text{row}}$ and $\psi_{\text{col}}$.
Monte Carlo simulations are used to obtain the results by running
at least $10^{6}$ Rayleigh flat fading channel realizations. The
channel state information at the receiver is considered to be perfectly
known.

\vspace{-3mm}

\subsection{Assessment of Expected Complexity for the RSD Algorithm}

The expression in (\ref{eq: RSD _expected comp numer}) is evaluated
for the two considered SM-MIMO systems using $\beta=7$. The expected
complexity coming from (\ref{eq: RSD _expected comp numer}) provides
almost identical results to (\ref{eq: RSD _expected comp}), however,
with added speed. The corresponding $w_{k}$ and $z_{k}$ at $\beta=7$
are given in {[}\ref{Ref: Ch7_19}, (Table 25.9){]}.

\begin{figure}
\begin{centering}
\includegraphics[scale=0.47]{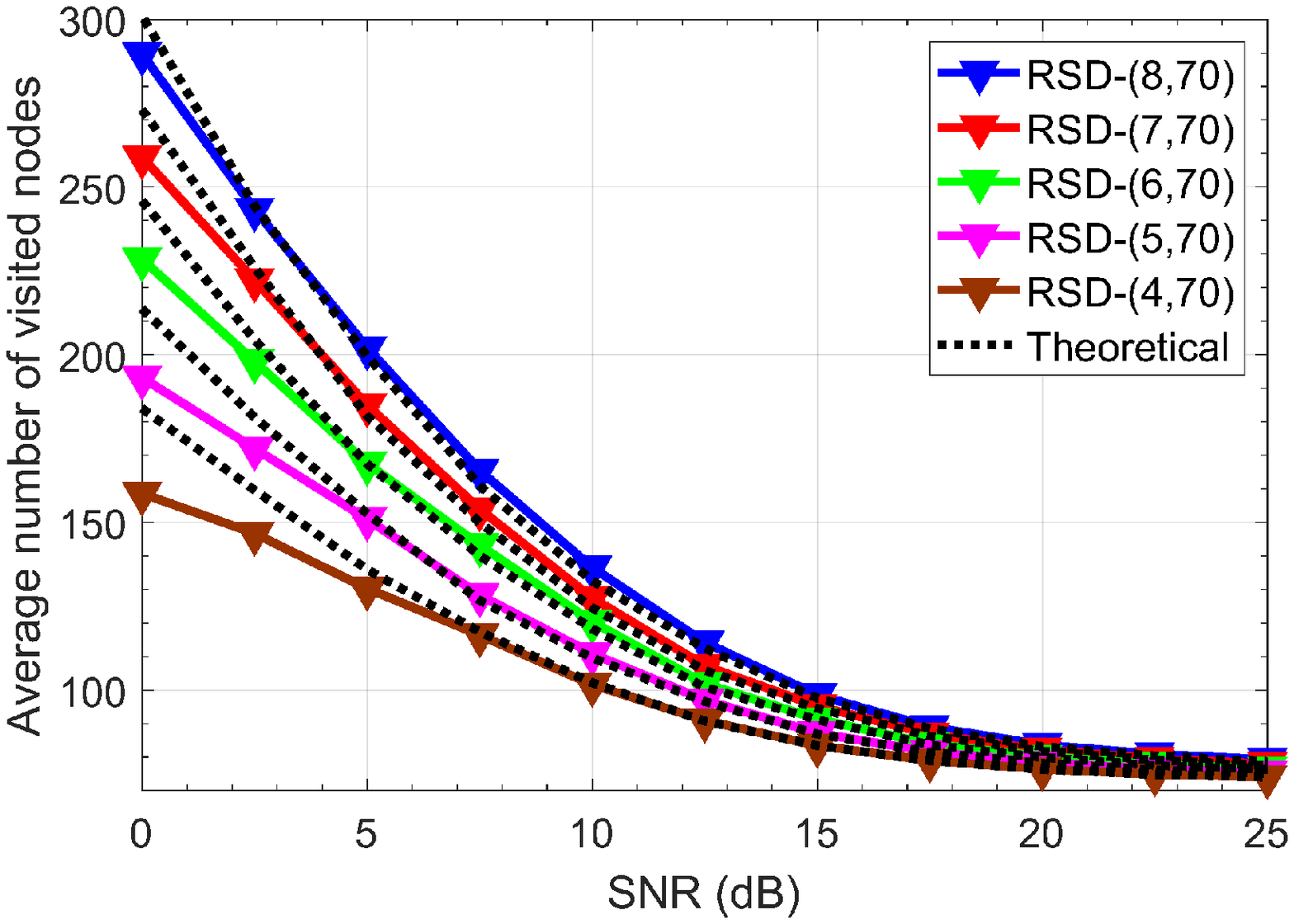}
\par\end{centering}
\caption{{\small{}\label{fig:ch7_Average-number-of 8x8}Average number of visited
nodes of the proposed RSD algorithm for $16$-QAM and $8\times8$
SM-MIMO system.}}
\end{figure}

\begin{figure}
\begin{centering}
\includegraphics[scale=0.47]{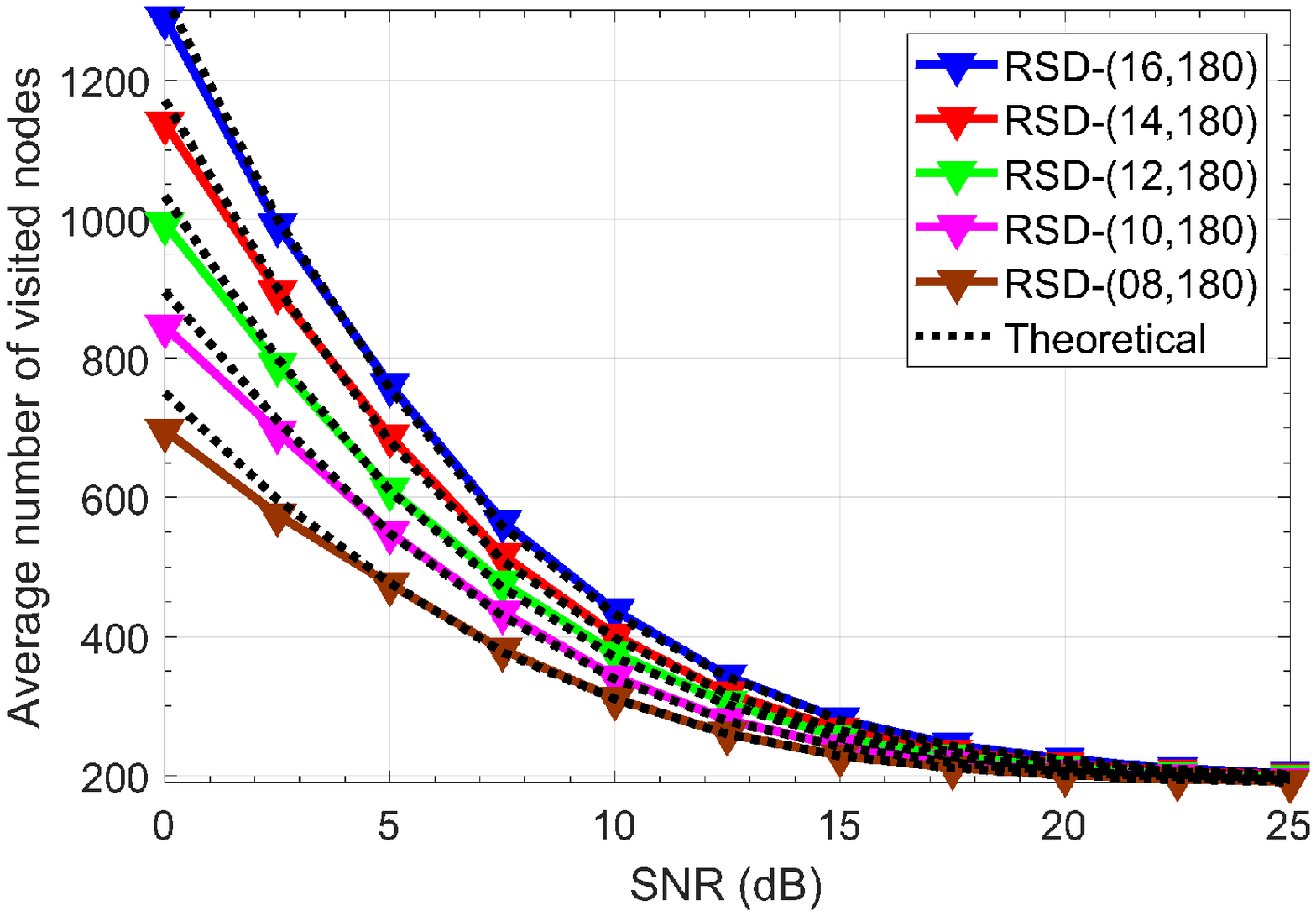}
\par\end{centering}
\caption{{\small{}\label{fig:ch7_Average-number-of 16x16}Average number of
visited nodes of the proposed RSD algorithm for $16$-QAM and $16\times16$
SM-MIMO system.}}
\end{figure}

Figures \ref{fig:ch7_Average-number-of 8x8} and \ref{fig:ch7_Average-number-of 16x16}
depict the average number of visited nodes of the RSD algorithm for
$16$-QAM with $8\times8$ SM-MIMO and $16$-QAM with $16\times16$
SM-MIMO, respectively. By decreasing $\psi_{\text{row}}$, the size
of the tree-search decreases and the complexity decreases correspondingly,
as shown in the figures. It is also notable that the RSD algorithm
requires less complexity to find the solution as the SNR increases.
As seen from these figures, the theoretical analysis in (\ref{eq: RSD _expected comp numer})
(or in (\ref{eq: RSD _expected comp})) provides a tight expression
for simulation results, for different values of $\psi_{\text{row}}$.
Note that (\ref{eq: RSD _expected comp numer}) perfectly matches
the simulation results in the higher SNR, which verifies the feasibility
of the pruned radius simplification assumption mentioned in Section
\ref{subsec:Expected-Complexity-Analysis}.

\vspace{-3mm}

\subsection{Comparisons with Literature Algorithms}

In this subsection, the BER and complexity are compared with those
of the literature algorithms (e.g., {[}\ref{Ref: Ch7_7}{]}, {[}\ref{Ref: Ch7_8}{]},
and {[}\ref{Ref: Ch7_10}{]}). The complexity comparison is assessed
by calculating the complexity reduction ratio which is defined as

\vspace{-1mm}

\begin{equation}
\Psi_{\text{Reduction}}^{\Omega}=\frac{MN_{t}N_{r}-\Psi^{\Omega}}{MN_{t}N_{r}}=1-\frac{\Psi^{\Omega}}{MN_{t}N_{r}},\label{eq: ch_6_Comp Reduction}
\end{equation}

\noindent where $\Psi_{\text{Reduction}}^{\Omega}$ is the complexity
reduction ratio for the $\Omega\in$ \{RSD, SD-{[}\ref{Ref: Ch7_7}{]},
SD-{[}\ref{Ref: Ch7_8}{]}, SD-{[}\ref{Ref: Ch7_10}{]}\} algorithm.

\begin{figure}
\begin{centering}
\includegraphics[scale=0.47]{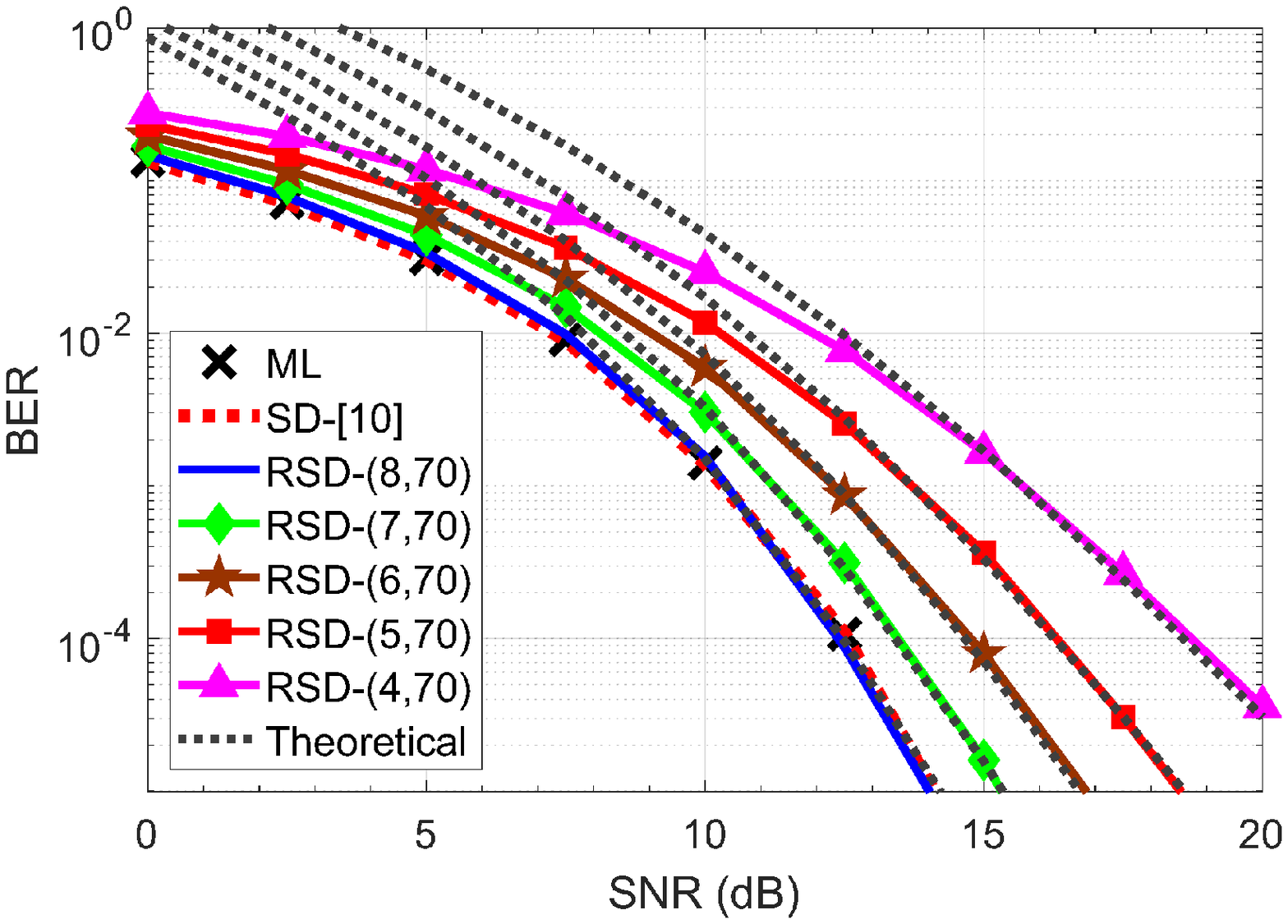}
\par\end{centering}
\caption{{\small{}\label{fig:ch_7_BER-performance-8x8}BER comparison for the
$16$-QAM and $8\times8$ SM-MIMO.}}
\end{figure}

\begin{figure}
\begin{centering}
\includegraphics[scale=0.47]{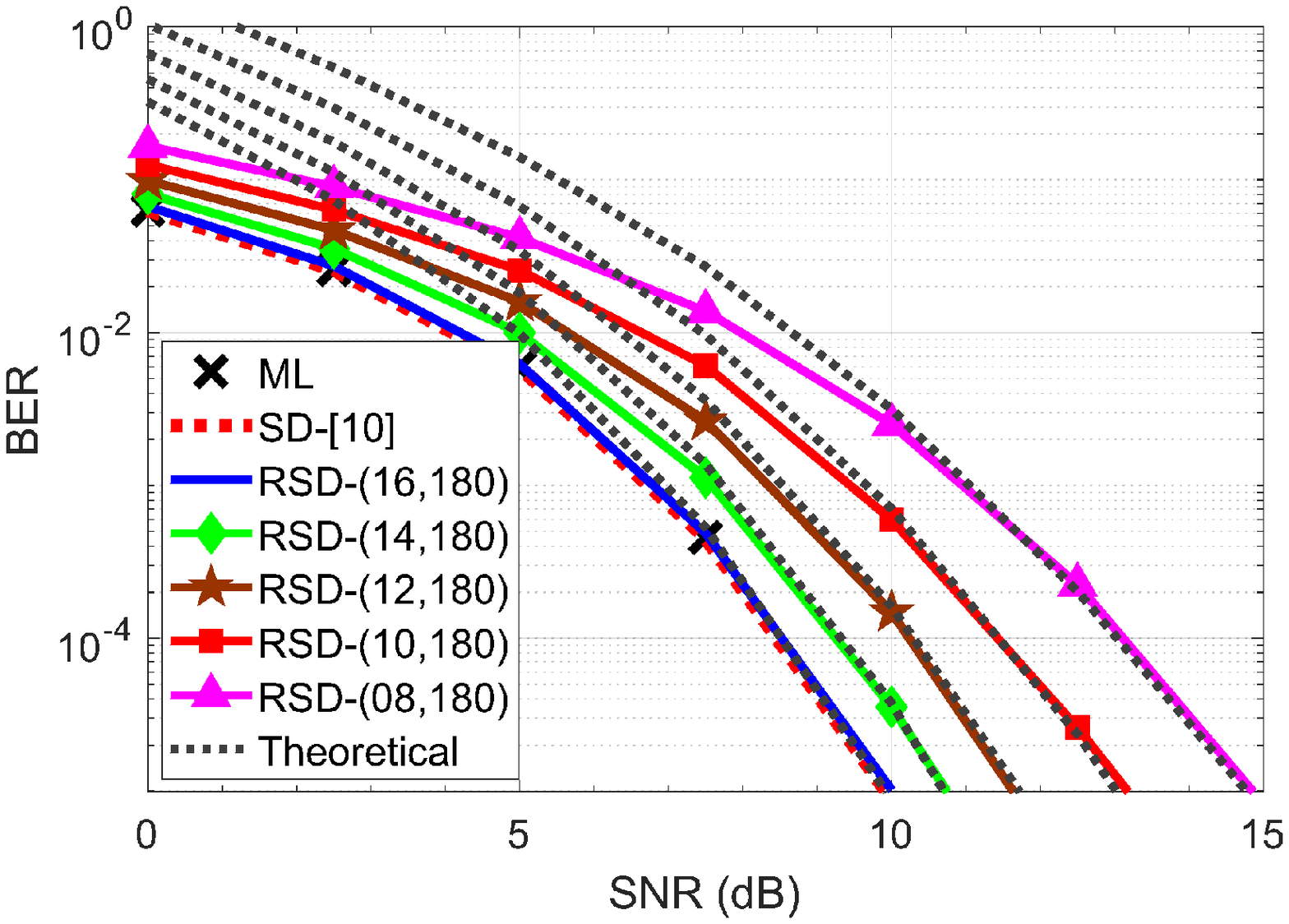}
\par\end{centering}
\caption{{\small{}\label{fig:ch_7_BER-performance-16x16}BER comparison for
the $16$-QAM and $16\times16$ SM-MIMO.}}
\end{figure}

Figures \ref{fig:ch_7_BER-performance-8x8} and \ref{fig:ch_7_BER-performance-16x16}
show the BER performance of the RSD algorithm compared to the optimum
algorithms, for $16$-QAM with $8\times8$ SM-MIMO and $16$-QAM with
$16\times16$ SM-MIMO, respectively. As shown from these figures,
the RSD-(8,70) and RSD-(16,180) provide the same BER as the ML BER
performance for $16$-QAM with $8\times8$ SM-MIMO and $16$-QAM with
$16\times16$ SM-MIMO, respectively. It should be noted that the SD-{[}\ref{Ref: Ch7_7}{]}
and SD-{[}\ref{Ref: Ch7_8}{]} algorithms provide the same BER performance
as the ML and SD-{[}\ref{Ref: Ch7_10}{]} algorithms, and their results
are omitted for the visibility of figures. Based on the reliable design
of the RSD algorithm, sub-optimal BER performances can be obtained
by varying the value of $\psi_{\text{row}}$. The BER analysis in
(\ref{eq: UB BER RSD}) is confirmed via simulation results.

\begin{figure}
\begin{centering}
\includegraphics[scale=0.47]{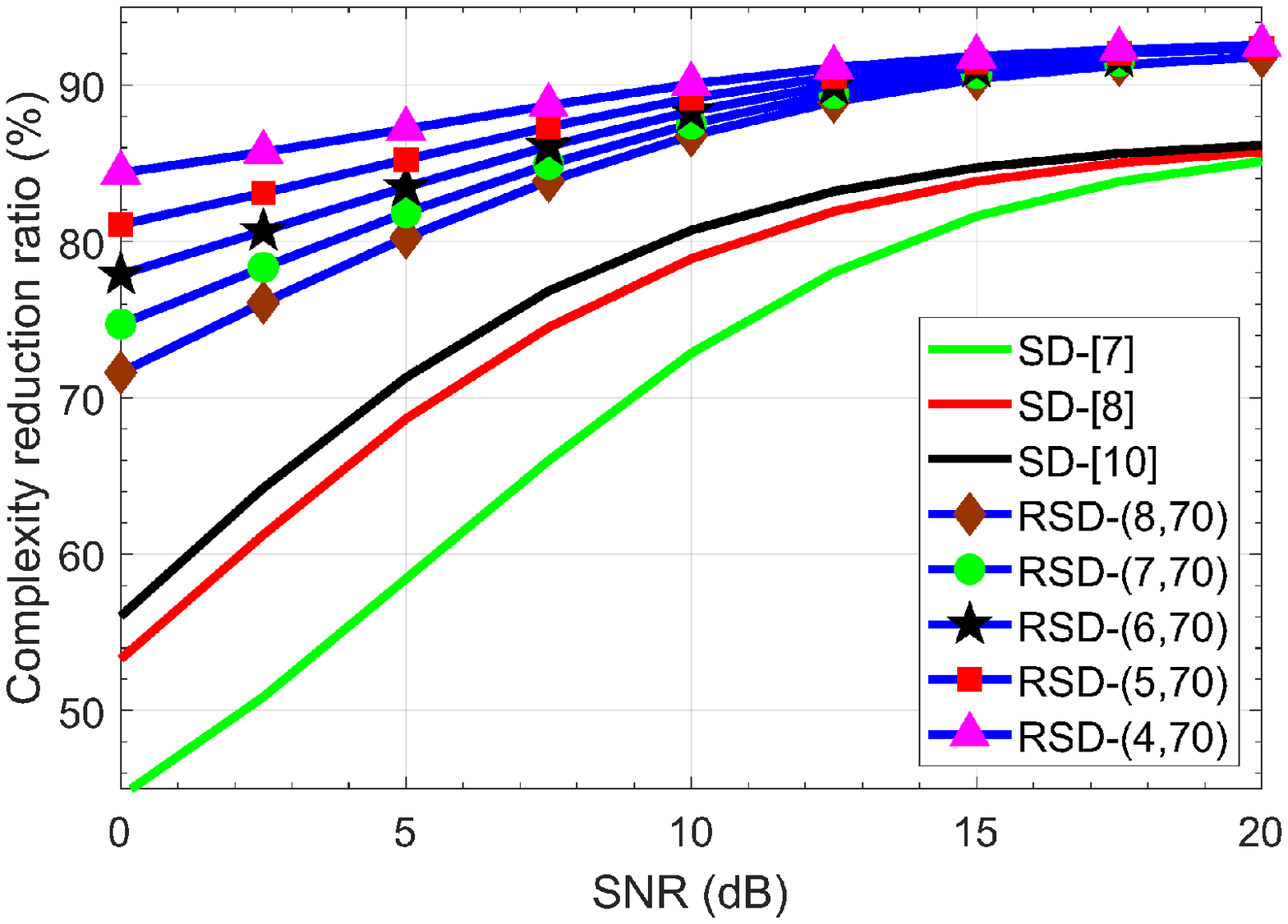}
\par\end{centering}
\caption{{\small{}\label{fig:ch_7_Complexity-reduction-8x8}Complexity reduction
comparison for the $16$-QAM and $8\times8$ SM-MIMO system.}}
\end{figure}

\begin{figure}
\begin{centering}
\includegraphics[scale=0.47]{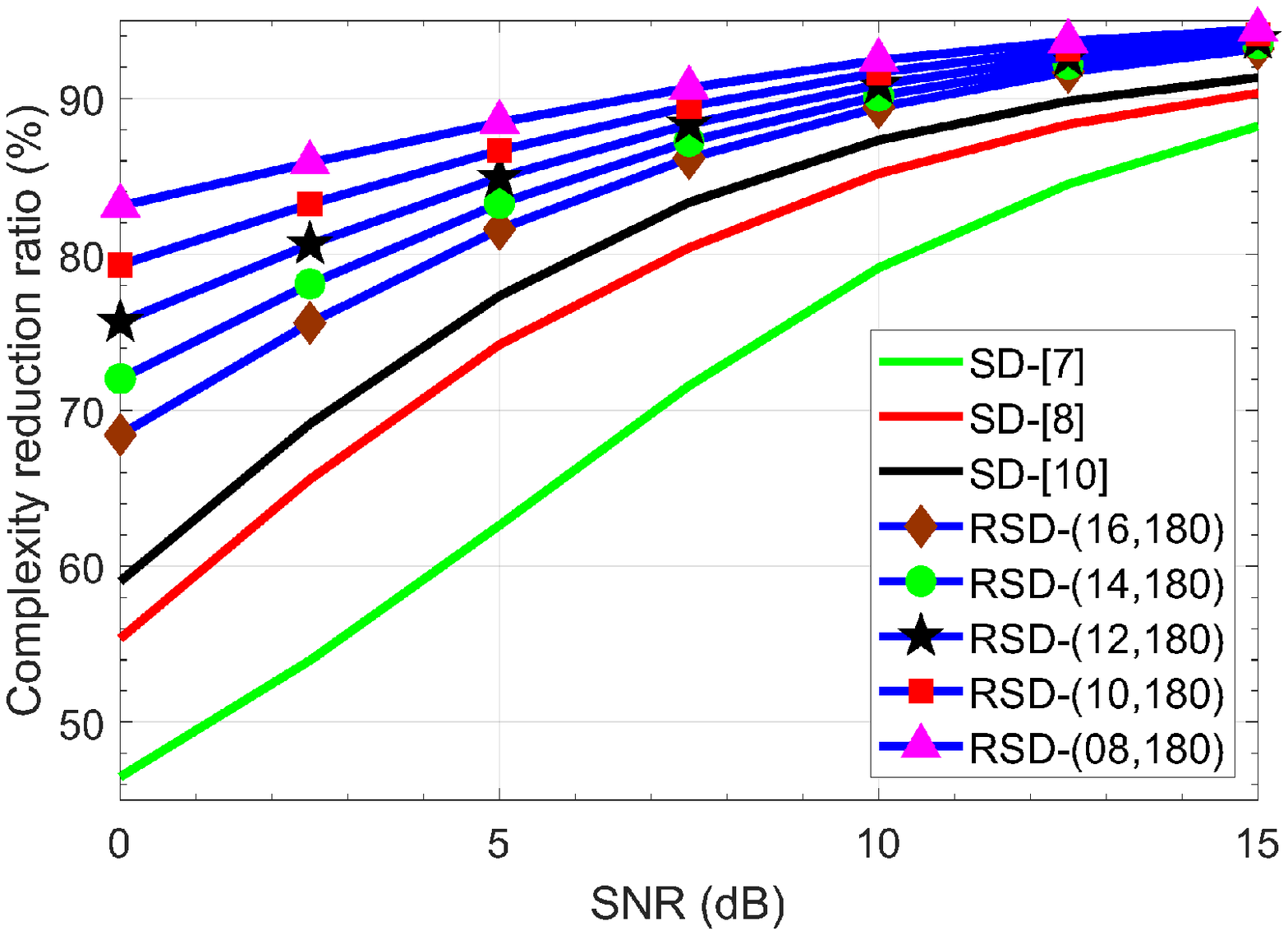}
\par\end{centering}
\caption{{\small{}\label{fig:ch_7_Complexity-reduction-16x16}Complexity reduction
comparison for the $16$-QAM and $16\times16$ SM-MIMO system.}}
\end{figure}

Figures \ref{fig:ch_7_Complexity-reduction-8x8} and \ref{fig:ch_7_Complexity-reduction-16x16}
depict the complexity reduction ratio of all algorithms for $16$-QAM
with $8\times8$ SM-MIMO and $16$-QAM with $16\times16$ SM-MIMO,
respectively. As seen from these figures, the RSD algorithm provides
the best reduction in complexity compared to all existing algorithms.
It also offers reliable decoding complexities that vary from $72\%$
to $92\%$ for $16$-QAM with $8\times8$ SM-MIMO and from $68\%$
to $95\%$ for $16$-QAM with $16\times16$ SM-MIMO. This reliability
in the decoding can fit a wide range of practical application requirements.

\section{Conclusion}

This paper proposes a novel reliable algorithm to decode SM transmitted
messages. The BER performance and complexity of the proposed algorithm
are theoretically derived. The proposed algorithm provides a significant
reduction in the decoding complexity (e.g., up to $95\%$) compared
to ML, without sacrificing the BER performance. A flexible trade-off
between the BER performance and complexity is presented to demonstrate
the reliability of the proposed algorithm.

\end{document}